\documentstyle[12pt]{article}
\begin{document}
\begin{center}
PARAMETER MISMATCHES, VARIABLE DELAY TIMES AND SYNCHRONIZATION IN 
TIME-DELAYED SYSTEMS\\
E. M. Shahverdiev \footnote {Regular Associate with the Abdus Salam ICTP;corresponding 
author:shahverdiev@physics.ab.az}, R.A.Nuriev\\
Institute of Physics, 370143 Baku,Azerbaijan\\
R.H.Hashimov\\
Azerbaijan Technical University, 370073 Baku,Azerbaijan\\
K. A. Shore\\
School of Informatics, University of Wales, Bangor, Dean Street, 
Bangor, LL57 1UT, Wales, UK\\
~\\
ABSTRACT\\
\end{center}
We investigate synchronization between two unidirectionally linearly coupled chaotic 
non-identical time-delayed systems and show that parameter mismatches are of crucial 
importance to achieve 
synchronization. We establish that independent of the relation between the delay time in 
the coupled systems and the coupling delay time, only retarded  synchronization with the 
coupling delay time is obtained. We show that with parameter mismatch or without it neither complete nor 
anticipating synchronization occurs. We derive existence and stability conditions for the 
retarded synchronization manifold. We demonstrate our approach using examples of the Ikeda 
and Mackey Glass models. Also for the first time we investigate chaos synchronization in 
time-delayed systems with variable delay time and find both existence and sufficient 
stability conditions for the retarded synchronization manifold with the coupling-delay 
lag time. Also for the first time we consider synchronization between two unidirectionally 
coupled chaotic multi-feedback Ikeda systems and derive existence and stability conditions 
for the different anticipating, lag, and complete synchronization regimes.\\
~\\
Key word(s):chaos synchronization, time-delayed systems, variable delay times, parameter 
mismatches, multi-feedback systems\\
~\\
\begin{center}
1.INTRODUCTION
\end{center}
\indent Seminal papers on chaos synchronization [1] have stimulated a wide range 
of research activity especially extensively in lasers, electronic circuits, chemical and 
biological systems [2]. Possible application areas of chaos synchronization are in 
secure communications, optimization of nonlinear system performance, modeling 
brain activity and pattern recognition phenomena [2].\\
\indent There are different types of sychronization in interacting chaotic systems. 
Complete, generalized, phase, lag and anticipating synchronizations of chaotic oscillators 
have been described theoretically and observed experimentally. Complete synchronization 
implies coincidence of states of interacting systems, $y(t)=x(t)$ [1]. Generalized 
synchronization is defined as the presence of some functional relation between the states 
of response and drive, i.e. $y(t)=F(x(t))$ [3]. Phase synchronization means entrainment of 
phases of chaotic oscillators, $n\Phi_{x}-m\Phi_{y}=const,$ ($n$ and $m$ are integers) 
whereas their 
amplitudes remain chaotic and uncorrelated [4].  Lag synchronization {\it for the first 
time} was introduced by Rosenblum {\it et al.}  [5]  under certain approximations in 
studying synchronization between {\it bi-directionally} coupled systems described by the 
ordinary differential equations (no intrinsic delay terms) with {\it parameter 
mismatches}:$y(t)\approx x_{\tau}(t)\equiv 
x(t- \tau)$ with positive $\tau$. Anticipating synchronization [6-8] also appears 
as a coincidence of shifted-in-time states of two coupled systems, but in this case the 
driven system anticipates the driver, $y(t)=x(t+\tau)$ or $x=y_{\tau}$,$\tau >0$. An 
experimental observation of anticipating synchronization in external cavity laser 
diodes [9]  has been reported recently, see also [10] for the theoretical interpretation 
of the experimental results.  The concept of inverse anticipating synchronization  
$x=-y_{\tau}$  is introduced in [11]. \\
\indent Due to finite signal transmission times, switching speeds and memory 
effects time-delayed systems are ubiquitous in nature, technology and society [12]. 
Therefore the study of synchronization phenomena in such systems is of high practical 
importance. Time-delayed systems are also interesting because the dimension of their 
chaotic dynamics can be made arbitrarily large by increasing their delay time. From 
this point of view these systems are especially appealing for secure communication 
schemes [13].\\
\indent  Role of parameter mismatches in synchronization phenomena is quite versatile. 
In certain cases parameter mismatches are detrimental to the synchronization quality: 
in the case of small parameter mismatches the synchronization error does not decay to 
zero with time, but can show small fluctuations about zero or even a non-zero mean 
value; larger values of parameter mismatches can result in the loss of synchronization 
[8,14]. In some cases parameter mismatches change the time shift between the synchronized 
systems [15]. In certain cases their presence is necessary for synchronization. We 
reiterate that the crucial role of parameter mismatches for lag synchronization 
between {\it bi-directionally} coupled systems  was first studied in [5] by  
Rosenblum {\it et al.}. As such, lag synchronization cannot be observed if two oscillators 
are completely identical, see e.g. [16] and references therein.\\
\indent Multi-feedback and multi-delay systems are ubiquitous in nature and technology. 
Prominent examples can be found in biological and biomedical systems,laser physics,
integrated communications [12]. In laser physics such a situation arises in lasers 
subject to two or more optical or elctro-optical feedback. Second optical feedback could 
be useful to stabilize laser intensity [17]. Chaotic behaviour of laser systems with two 
optical feedback mechanism is studied in recent works [18]. To the best of our knowledge 
chaos synchronization between the multi-feedback systems is to be investigated yet. Having 
in mind enormous application implications of chaos synchronization e.g. in secure 
communication, investigation of synchronization regimes  (lag, complete, anticipating 
etc.) in multi-feedback systems is of immense  importance.\\
\indent In this paper we investigate synchronization between the two {\it 
unidirectionally} coupled chaotic non-identical time-delayed systems having a fairly 
general form of coupling and show for the first time that parameter mismatches are, in 
fact, of crucial importance for achieving synchronization. We show that independent of 
the relation between the delay time in the coupled systems and the coupling delay time, 
only retarded (lag) synchronization is obtained. ( Usually for lag synchronization 
between the unidirecitionally coupled  time-delayed systems the term retarded 
synchronization is prefered [8].) In this case the lag time is the coupling delay 
time. We consider both constant and variable feedback delay times. We demonstrate our 
approach using examples of the Ikeda and Mackey Glass models. \\
\indent  In the paper also for the first time we investigate synchronization between two 
unidirectionally coupled chaotic multi-feedback Ikeda systems and find both existence and 
stability conditions for different synchronization regimes.\\
\begin{center}
2.GENERAL THEORY
\end{center}
\indent Consider a situation where a time-delayed chaotic master (driver) system 
$$\hspace*{6cm}\frac{dx}{dt}=-\alpha_{1} x + 
k_{1}f(x_{\tau_{1}}),\hspace*{6.5cm}(1)$$
drives a non-identical slave (response) system
$$\hspace*{6cm}\frac{dy}{dt}=-\alpha_{2} y + k_{2} 
f(y_{\tau_{1}})+k_{3}x_{\tau_{2}},\hspace*{5.2cm}(2)$$
where $x$ and $y$ are dynamical variables; $f(x)$ is differentiable nonlinear 
function; $\alpha_{1}$ and $\alpha_{2}$ are 
relaxation coefficients for the driving and driven dynamical variables, 
respectively:throughout the paper we assume that 
$\alpha_{1}=\alpha - \delta$ and $\alpha_{2}=\alpha + \delta$, $\delta$ determines 
the mismatch of relaxation coefficents; 
$\tau_{1}$ is the feedback delay time in the coupled systems;$\tau_{2}$ is the 
coupling delay time between the systems. $k_{1}$ 
and $k_{2}$ are the feedback rates for the master and the response systems, 
respectively;$k_{3}$ is the linear coupling rate 
between the driver and the response system.\\
Now we will show that chaotic systems (1) and (2) can be synchronized on the 
retarded synchronization manifold with the lag time $\tau_{2}$: 
$$\hspace*{7cm}y=x_{\tau_{2}}.\hspace*{8.2cm}(3)$$
We denote the error signal by $\Delta=x_{\tau_{2}}-y$. Then from systems (1) and 
(2) we find the following error dynamics:
$\frac{d\Delta}{dt}=-\alpha_{2}\Delta + (2\delta -k_{3})x_{\tau_{2}}+ 
k_{1}f(x_{\tau_{1}+\tau_{2}})-k_{2}f(y_{\tau_{1}})$.
Thus under conditions
$$\hspace*{6cm}2\delta =k_{3}, k_{1}=k_{2},\hspace*{7.8cm}(4)$$ 
the error dynanics can be written as:
$$\hspace*{5cm}\frac{d\Delta}{dt}=-\alpha_{2}\Delta + k_{1} 
\Delta_{\tau_{1}}f^{\prime}(x_{\tau_{1}+\tau_{2}}).\hspace*{6.1cm}(5)$$
It is obvious that $\Delta=0$ is a solution of system (5). To study the sufficient 
stability condition for the retarded synchronization manifold $y=x_{\tau_{2}}$ one can 
use a Krasovskii-Lyapunov functional approach [12, 19].\\ 
The sufficient stability condition for the trivial solution $\Delta=0$ of eq.(5) can be 
found by investigating the positively defined Krasovskii-Lyapunov fuctional 
$$\hspace*{5cm}V(t)=\frac{1}{2}\Delta^{2} + 
\mu\int_{-\tau}^{0}\Delta^{2}(t+t_{1})dt_{1},\hspace*{5.5cm}(6)$$ 
where $\mu >0$ is an arbitrary positive parameter. According to [12,19], the solution 
$\Delta=0$ is stable, if the derivative of the 
functional (6) along the trajectory of equation $\frac{d\Delta}{dt}=-r(t)\Delta - 
s(t)\Delta_{\tau}$ is negative. In general this negativity condition is of the form:
$4(r-\mu)\mu>s^{2}$ and $r>\mu>0$. As the value of $\mu$ that will allow $s^{2}$ as 
large as possible is $\mu=\frac{r}{2}$, the asymptotic stability condition for $\Delta=0$ 
can be written as 
$$\hspace*{7cm}r^{2}>s^{2},\hspace*{8.2cm}(7)$$ 
which is equivalent to $r>\vert s \vert $. This result is valid for both constant and 
time-dependent coefficients $r$ and $s$ (in the latter case $r(t)$ and $s(t)$ should be 
bounded continuous functions [12]). Thus we obtain that 
$$\hspace*{5cm}\alpha_{2} > \vert k_{1}f^{\prime}(x_{\tau_{1}+\tau_{2}}) \vert 
\hspace*{8.6cm}(8)$$
is the sufficient stability condition for retarded synchronization manifold (3). The 
condition (4) is the existence condition of retarded synchronization between the 
unidirectionally coupled systems (1) and (2).\\
Thus we find that under certain conditions systems (1) and (2) admit the retarded chaos 
synchronization manifold $y=x_{\tau_{2}}$ {\it only} under parameter mismatch ie  
$\alpha_{1}\neq \alpha_{2}$. We also notice that without the parameter mismatch, i.e. 
$\alpha_{1}=\alpha_{2}=\alpha$ neither 
$y=x_{\tau_{2}-\tau_{1}}$ nor $y=x_{\tau_{1}-\tau_{2}}$ is the  synchronization manifold. 
We also emphasize that, in general for both 
$\alpha_{1}=\alpha_{2}$ and $\alpha_{1}\neq \alpha_{2}$ systems (1) and (2) admits neither 
complete nor anticipating {\it chaos} synchronization.\\ 
\indent So far we have considered the case of constant feedback delay time 
$\tau_{1}$. It is of immense  interest to study chaos synchronization in time-delayed 
systems with variable feedback delay time. Basic interest is driven by the fact that so 
far there are no reported research on this particular subject in the literature. 
Practical interest is motivated by the appreciation that time-delayed systems with 
variable delay times are more realistic. As an example one can refer to the biological 
biorhythms, where the capacity of assimilation of nutrients by an organism varies cyclicly 
during the day [20] .\\
Now we will try to find both the existence and stability conditions for the synchronization 
manifold (3) in the case of variable feedback delay times. It is straightforward to 
establish that the analog of the error  dynamics equation in the case of variable delay 
time $\tau_{1}(t)$ is of the form:
$$\hspace*{5cm}\frac{d\Delta}{dt}=-\alpha_{2}\Delta + k_{1} 
\Delta_{\tau_{1}(t)}f^{\prime}(x_{\tau_{1}(t)+\tau_{2}}).\hspace*{5.7cm}(9)$$
Again, as in the case of constant feedback delay times equation (9) is obtained 
from the studying the coevolution of eqs.(1) and (2) along the manifold (3). Analysis 
of the error dynamics shows that the existence conditions (4) hold for the variable delay 
cases. Next let us find the sufficient stability condition for system (9). According to 
[12] for that purpose one can still use the functional (6). Namely as presented in [12], 
when $\tau=\tau 
(t)$ is continuously differentiable and bounded, the  solution $\Delta=0$ to 
$\frac{d\Delta}{dt}=-r(t)\Delta -s(t)\Delta_{\tau (t)}$ is 
uniformly asymptotically stable, if $a(t)>\mu>0$ and 
$(2r(t)-\mu)(1-\frac{d\tau}{dt})\mu>s^{2}(t)$ uniformly in $t$. Applying the same 
procedure as in the case of constant feedback delay time, we can find the value of 
$\mu$ that will allow $s^{2}$  to be as large as possible: $\mu=r$. Thus we find that 
the sufficent stability condition for the $\Delta=0$ solution of time delay equation with 
time dependent coefficients $\frac{d\Delta}{dt}=-r(t)\Delta 
-s(t)\Delta_{\tau (t)}$ is:
$$\hspace*{5cm}r^{2}(t)(1-\frac{d\tau (t)}{dt})>s^{2}(t).\hspace*{7cm}(10)$$
Notice that for the constant delay time cases the inequality (10) is reduced to the 
well-known sufficent stability condition $r>\vert s \vert $.\\
As in our case $r=\alpha_{2}$ and $s=-k_{1}f^{\prime}(x_{\tau_{1}(t)+\tau_{2}})$
then the sufficent stability condition for synchronization manifold (3) for the 
time-delayed equations (1) and (2) with time dependent feedback delay $\tau_{1}$ can be 
written as:
$$\hspace*{4cm}\alpha_{2}^{2}(1-\frac{d\tau_{1}(t)}{dt}) > 
(k_{1}f^{\prime}(x_{\tau_{1}(t)+\tau_{2}}))^{2}.\hspace*{6.3cm}(11)$$
\begin{center}
3.1.EXAMPLE 1:THE IKEDA MODEL\\
\end{center} 
\indent In this subsection we demonstrate our general theory using the example of the 
Ikeda model. This investigation is of considerable practical importance, as the equations 
of the class B lasers with feedback (typical representatives of class B are solid-state, 
semiconductor, and low pressure $CO_{2}$ lasers [21]) can be reduced to an equation of 
the Ikeda type [22]. 
Consider synchronization between the Ikeda systems [6],
$$\hspace*{-3.2cm}\frac{dx}{dt}=-\alpha_{1} x - \beta \sin x_{\tau_{1}},$$
$$\hspace*{5cm}\frac{dy}{dt}=-\alpha_{2} y - \beta \sin y_{\tau_{1}} + 
Kx_{\tau_{2}}.\hspace*{6.2cm}(12)$$
The Ikeda model was introduced to describe the dynamics of an optical bistable 
resonator and is well-known for delay-induced chaotic behavior [23]. Physically $x$ is 
the phase lag of the electric field across the resonator; $\alpha$ is the relaxation 
coefficient; $\beta$ is the laser intensity injected into the system. $\tau_{1}$ is the 
round trip time of the light in the resonator or feedback delay time in the coupled 
systems;$\tau_{2}$ is the coupling delay time between systems $x$ and $y$.\\
First we consider the case of constant feedback delay time and show that 
$y=x_{\tau_{2}}$ is the retarded synchronization manifold, if the 
parameter mismatch $\alpha_{2}-\alpha_{1}=2\delta $ is equal to the coupling rate 
$K$. This can be seen by the dynamics of the error $\Delta=x_{\tau_{2}}-y$:
$$\hspace*{5cm}\frac{d\Delta}{dt}=-(\alpha + \delta)\Delta + (2\delta-K)x_{\tau_{2}} 
-\beta \cos x_{\tau_{1}+\tau_{2}}\Delta_{\tau_{1}}.\hspace*{2.7cm}(13)$$
(As in this example under study we choose feedback rates ($\beta $) equal for both 
the driver and driven systems, the second of the existence conditions in (4) becomes 
redundant.) The sufficient stability condition for the retarded synchronization manifold 
$y=x_{\tau_{2}}$ can be written as:$\alpha +\delta =\alpha_{2}>\vert\beta\vert $. Thus, 
as in case of general approach, we find that the retarded chaos 
synchronization manifold $y=x_{\tau_{2}}$ occurs {\it only} under 
parameter mismatch ie  $\alpha_{1}\neq \alpha_{2}$. By analyzing 
the corresponding error dynamics one can also establish that without the 
parameter mismatch, i.e. $\alpha_{1}=\alpha_{2}=\alpha$ neither 
$y=x_{\tau_{2}-\tau_{1}}$ nor $y=x_{\tau_{1}-\tau_{2}}$ is the synchronization manifold. 
We also emphasize that  for both $\alpha_{1}=\alpha_{2}$ and $\alpha_{1}\neq \alpha_{2}$ 
system  (12)  admits  neither complete (we notice that for special case of $\tau_{2}=0$ 
$y=x_{\tau_{2}}$ is the complete synchronization manifold, which exists if  
$\alpha_{1}\neq \alpha_{2}$) 
 nor anticipating {\it chaos} synchronization. We emphasize that this result is due to the 
linear coupling between the synchronized systems. The importance of the role of the form 
of coupling  between the synchronized systems is underlined in [6,24]. In the case of 
nonlinear (sinusoidal) coupling  for  identical drive and response Ikeda  systems, 
depending on the relation  between the feedback delay time and  the coupling delay 
time retarded, complete or anticipating  synchronization can occur,  see, e.g. [25] 
and references therein.\\
According to estimations the parameters values   
$\alpha_{1}=5$, $\alpha_{2}=25$, $\beta =20$, $K=20$ and $\tau_{1}=1$, $\tau_{2}=2$ (or 
 $\tau_{1}=3$, $\tau_{2}=1$) satisfy both existence and stability conditions for the 
retarded chaos synchronizartion manifold $y=x_{\tau_{2}}$ for the coupled systems (12).\\
{\indent} Next we consider the case of time dependent delay time $\tau_{1}(t)$.
First we notice that as in the case of time-independent delay times 
$2\delta =K$ is the condition of existence for the $y=x_{\tau_{2}}$ synchronization 
manifold. Next applying the general formula (11) derived earlier in the paper we write 
the sufficent stability condition for the synchronization manifold $y=x_{\tau_{2}}$ in 
the following form:
$$\hspace*{6cm}\alpha_{2}^{2}(1-\frac{d\tau_{1}(t)}{dt}) > \beta 
^{2},\hspace*{6.7cm}(14)$$
As an example consider 
the following sinusoidal form of the variable delay time :
$$\hspace*{6cm}\tau _{1} (t)=\tau_{0}+\tau_{a}\sin(\omega t),\hspace*{6.4cm}(15)$$
where $\tau_{0}$ is the zero frequency component;$\tau_{a}$ is the amplitude;
$\frac{\omega}{2\pi}$ is the frequency of the modulation. Then for the concrete form 
of variable delay time (15) the sufficient stability condition (14) can be writen as:
$$\hspace*{5cm}\alpha_{2}^{2}(1-
\tau_{a}\omega\cos(\omega t))>\beta^{2}.\hspace*{7cm}(16)$$
\begin{center}
3.2.EXAMPLE 2:THE MACKEY GLASS MODEL\\
\end{center}
In this subsection we demonstrate our approach using the example of the Mackey Glass 
model. The Mackey Glass model has been introduced as a model of blood generation for 
patients with leukemia and nowadays is very popular in chaos theory [19]. \\
Consider synchronization between the Mackey Glass systems
$$\hspace*{-3.3cm}\frac{dx}{dt}=-\alpha_{1} x+k_{1} \frac{a_{1}x_{\tau_{1}}}{1+x_{\tau_{1}}^{b}} ,$$
$$\hspace*{4.5cm}\frac{dy}{dt}=-\alpha_{2} y +k_{2} \frac{a_{2}x_{\tau_{1}}}{1+x_{\tau_{1}}^{b}}
+ k_{3}x_{\tau_{2}}.\hspace*{6.2cm}(17)$$
The dynamical variable in the Mackey Glass model is the concentration of the mature cells 
in blood at time $t$ and the delay time is the time between the initiation of cellular 
production in the bone marrow and the release of mature cells into the blood [20].\\
Again by investigating the corresponding error dynamics we can show that 
$y=x_{\tau_{2}}$ is the retarded synchronization manifold, if the parameter mismatch 
$\alpha_{2}-\alpha_{1}=2\delta $ is equal to the coupling rate $k_{3}$ and 
$k_{1}a_{1}=k_{2}a_{2}$. We notice that here we can allow for parameter mismatches 
for $a$,
and thus have more flexibility to achieve synchronization. With these existence 
conditions, the sufficient stability condition for the retarded synchronization 
manifold 
$y=x_{\tau_{2}}$ can be written as:
$\alpha_{2} > \vert k_{1}a_{1}f^{\prime}(x_{\tau_{1}+\tau_{2}}) \vert ,$ with  $f(x_{\tau})=\frac{x_{\tau}}{1+x_{\tau}^{b}}$.\\
For analytical estimation of $\alpha_{2}$ we take into account that the 
absolute maximum of the function $ \vert f^{\prime} (x_{\tau}) \vert $  is obtained
at $x_{\tau}=(\frac{b+1}{b-1})^{\frac{1}{b}}$ and is equal to $ 
\frac{(b-1)^{2}}{4b}$ [19]. Thus we arrive at the following sufficient stability 
condition for the 
synchronization manifold $y=x_{\tau_{2}}$ for the coupled systems (17) :
$$\hspace*{5.4cm}\alpha_{2} > k_{1}a_{1}\frac{(b-1)^{2}}{4b}.\hspace*{7.8cm}(18)$$
Again we would like to underline that only {\it retarded} synchronization occurs 
notwithstanding the relation between the feedback delay time and coupling delay time; 
moreover  for both 
$\alpha_{1}=\alpha_{2}$ and $\alpha_{1}\neq \alpha_{2}$ coupled systems (17) admit 
neither complete nor anticipating {\it chaos} synchronization.\\
According to estimations the parameters values   $\alpha_{1}=0.1$, $\alpha_{2}=5$, 
$k_{1}a_{1}=2$,$b =10$, $k_{3}=4.9$,$\tau_{1}=10$ and $\tau_{2}=20$ 
(or $\tau_{1}=50$ and $\tau_{2}=20$, $\tau_{1}=\tau_{2}=50$) satisfy both existence and 
stability conditions for the retarded chaos synchronizartion manifold $y=x_{\tau_{2}}$ 
for the coupled systems (17).\\
\begin{center}
4.SYNCHRONIZATION BETWEEN THE MULTI-FEEDBACK IKEDA SYSTEMS\\
\end{center}
\indent Consider synchronization between the multi-feedback Ikeda systems,
$$\hspace*{5cm}\frac{dx}{dt}=-\alpha x + m_{1} \sin x_{\tau_{1}}
+m_{2} \sin x_{\tau_{2}},\hspace*{4.6cm}(19)$$
$$\hspace*{5cm}\frac{dy}{dt}=-\alpha y + m_{3} \sin y_{\tau_{1}}
+ m_{4} \sin y_{\tau_{2}} + K \sin x_{\tau_{3}},\hspace*{2.9cm}(20)$$
with positive $\alpha_{1,2}$ and $-m_{1,2,3,4}$.\\ 
As mentioned above physically $x$ is the phase lag of the electric field across the 
resonator; $\alpha$ is the relaxation coefficient for the driving $x$ and driven $y$ 
dynamical variables;$-m_{1,2}$ and $-m_{3,4}$ are the laser intensities injected into 
the driving and driven systems,respectively. $\tau_{1,2}$ are the feedback delay times 
in the coupled systems;$\tau_{3}$ is the coupling delay time between systems $x$ and 
$y$;$K$ is the coupling rate between  the driver $x$ and the response system $y$.\\
First we will show that systems (19) and (20) can be synchronized on the manifold:
$$\hspace*{7cm}y=x_{\tau_{3}-\tau_{1}}.\hspace*{7.7cm}(21)$$
(For $\tau_{3}>\tau_{1}$,$\tau_{3}=\tau_{1}$, and $\tau_{3}<\tau_{1}$ 
 (21) is the retarded, complete, and anticipating synchronization manifold, respectively.)  
We denote the error signal by $\Delta=x_{\tau_{3}-\tau_{1}}-y$. Then from  systems (19) 
and (20)we find the following error dynamics 
$\Delta=x_{\tau_{3}-\tau_{1}}-y$:$\frac{d\Delta}{dt}=-\alpha\Delta +
((m_{1}-K)\sin x_{\tau_{3}}-m_{3}\sin y_{\tau_{1}}) + m_{2}\sin x_{\tau_{2}+\tau_{3}-\tau_{1}}-m_{4}\sin y_{\tau_{2}}$ 
Thus under conditions
$$\hspace*{6cm}m_{1}-K=m_{3}, m_{2}=m_{4}\hspace*{6.3cm}(22)$$
the error dynamics can be written as:
$$\hspace*{5cm}\frac{d\Delta}{dt}= -\alpha\Delta + m_{3} \Delta_{\tau_{1}} \cos x_{\tau_{3}} + m_{2} \Delta_{\tau_{2}} \cos  x_{\tau_{2}+\tau_{3}-\tau_{1}}.\hspace*{2.6cm}(23)$$
It is obvious that $\Delta= 0$ is the solution of system (23).To study the stability of 
the synchronization manifold $y=x_{\tau_{3}-\tau_{1}}$ one can again use a 
Krasovskii-Lyapunov functional approach. According to [12], the sufficient stability 
condition for the trivial solution $\Delta=0$ of time-delayed equation 
$\frac{d\Delta}{dt}=-r(t)\Delta + s_{1}(t)\Delta_{\tau_{1}}+ s_{2}(t)\Delta_{\tau_{2}} $ 
is: $r(t)>\vert s_{1}(t) \vert + \vert s_{2}(t) \vert $.\\
Thus we obtain that the sufficient stability condition for the synchronization manifold 
$y=x_{\tau_{3}-\tau_{1}}$ (21) can be written as:
$$\hspace*{5cm}\alpha > \vert m_{3} \vert +\vert m_{2} \vert.\hspace*{9cm}(24)$$ 
Conditions (22) are the existence conditions for the synchronization manifold (21) 
between the unidirectionally coupled multi-feedback systems (19) and (20).\\
Analogously we find that $y=x_{\tau_{3}-\tau_{2}}$ is the synchronization manifold  
between systems (19) and (20) with corresponding existence $m_{2}-K=m_{4}$ and 
$m_{1}=m_{3}$  and stability conditions $\alpha > \vert m_{3} \vert + \vert m_{4} \vert $.\\
\indent One can easily generalize the previous rezults to $n$-tuple feedback Ikeda 
systems. Indeed consider synchronization between the following Ikeda models:\\
$$\hspace*{1cm}\frac{dx}{dt}=-\alpha x +m_{1x} \sin x_{\tau_{1}} 
+ m_{2x} \sin x_{\tau_{2}}+ \cdots +m_{nx}\sin x_{\tau_{n}},\hspace*{5.3cm}(25)$$
$$\hspace*{2cm}\frac{dy}{dt}=-\alpha y +m_{1y} \sin y_{\tau_{1}}
+m_{2y} \sin y_{\tau_{2}} +m_{ny} \sin y_{\tau_{n}}+ k \sin x_{\tau_{k}},\hspace*{3.4cm}(26)$$
Then the existence and sufficent stability conditions e.g. for the 
synchronization manifold $y=x_{\tau_{k}-\tau_{1}}$ are:$m_{1x}-k=m_{1y},m_{nx}=m_{ny}$ 
and $\alpha > \vert m_{1y} \vert + \vert m_{2y}\vert + \cdots + \vert m_{ny} \vert $,
respectively. For synchronization manifold  $y=x_{\tau_{k}-\tau_{2}}$, $m_{2x}-k=m_{2y}$ 
and $m_{nx}=m_{ny}$ are 
the existence conditions, and  $\alpha >\vert m_{1y} \vert +\vert m_{2y} \vert + 
\cdots + \vert m_{ny} \vert$ is the sufficient stability condition.\\
\begin{center}
5.CONCLUSIONS\\
\end{center}
\indent In this paper we have studied the relation between parameter  mismatches and 
synchronization in a certain class of  unidirectionally linearly coupled time-delayed 
systems and have shown for the first time that parameter  mismatches are of crucial 
importance for achieving synchronization. We have showed that independent of the relation 
between the feedback delay time in the coupled systems and the coupling delay time, only 
retarded (lag) synchronization with coupling delay lag time is obtained. We have 
established that either with parameter mismatch or without it neither complete nor 
anticipating chaos synchronization occurs.  We have demonstrated our approach using the 
Ikeda and Mackey Glass models. We mention that, for example in the case of  nonlinear 
(sinusoidal) coupling  for  identical drive and response Ikeda  systems, depending on 
the relation  between the feedback delay time and  the coupling delay time  retarded, 
complete or anticipating  synchronization can occur [25]. These results are of  
significant interest in the context of relationship between parameter mismatches, 
coupling forms and synchronization. Indeed, having in mind possible practical 
applications of anticipating chaos synchronization [6] in secure communications 
(anticipation of the future states of the transmitter (master laser) at the receiver 
(slave laser) allows more time to decode the message ), in the control of delay-induced 
instabilites in a wide range of non-linear systems, for the understanding of natural 
information processing choosing the ``appropriate"  parameters' mismatches and coupling 
forms certain types of synchronization can be switched off/on. We have also {\it for the 
first time} investigated chaos synchronization in variable delay time systems and found 
both existence and sufficient stability conditions for the retarded synchronization 
manifold with the coupling-delay lag time.\\
Also for the first time we have investigated synchronization 
between two unidirectionally coupled multi-feedback systems. These findings are of 
considerable interest in the context of synchronization between the stabilized laser 
systems (arrays) which hold great promise for space communication applications,where 
compact sources with high intensities are required. Additionally, synchronization 
between the multi-feedback systems can provide more flexibility in practical 
applications.\\
~\\
5.ACKNOWLEDGEMENTS\\
~\\
Concluding parts of the work have been done at the Abdus Salam ICTP. E.M.Shahverdiev 
kindly acknowledges a very helpful discussions with Professor H.A.Cerdeira.
E.M. Shahverdiev also acknowledges support from the UK Engineering and Physical Sciences 
Research Council grant GR/R22568/01 and The Abdus Salam ICTP Associate scheme.\\

\end{document}